\documentclass[a4paper,fleqn,usenatbib]{mnras}
\usepackage{newtxtext,newtxmath}

\usepackage[T1]{fontenc}
\usepackage{ae,aecompl}

\usepackage{graphicx}	
\usepackage{amsmath}	
\usepackage{amssymb}	

\title[The RM effect of ringed planets]{Characterising exo-ringsystems around fast-rotating stars using the Rossiter-McLaughlin effect}

\author[E.J.W. de Mooij et al.]{
Ernst J.W. de Mooij,$^{1,2}$\thanks{E-mail: ernst.demooij@dcu.ie (EdM)}
Christopher A. Watson,$^{2}$
Matthew A. Kenworthy$^{3}$
\\
$^{1}$School of Physical Sciences, Dublin City University, Glasnevin, Dublin 9, Ireland\\
$^{2}$Astrophysics Research Centre, Queen's University Belfast, Belfast BT7 1NN, United Kingdom\\
$^{3}$Leiden Observatory, Leiden University, Niels Bohrweg 2, 2333 RA Leiden, The Netherlands
}

\date{Accepted XXX. Received YYY; in original form ZZZ}

\pubyear{2017}

\begin{document}
\label{firstpage}
\pagerange{\pageref{firstpage}--\pageref{lastpage}}
\maketitle

\begin{abstract}
  Planetary rings produce a distinct shape distortion in transit lightcurves.
  However, to accurately model such lightcurves the observations need to cover
  the entire transit, especially ingress and egress, as well as an
  out-of-transit baseline. Such observations can be challenging for long period
  planets, where the transits may last for over a day. Planetary rings will also
  impact the shape of absorption lines in the stellar spectrum, as the planet
  and rings cover different parts of the rotating star (the Rossiter-McLaughlin
  effect). These line-profile distortions depend on the size, structure,
  opacity, obliquity and sky projected angle of the ring system. For slow
  rotating stars, this mainly impacts the amplitude of the induced velocity
  shift, however, for fast rotating stars the large velocity gradient across the
  star allows the line distortion to be resolved, enabling direct determination
  of the ring parameters. We demonstrate that by modeling these distortions we
  can recover ring system parameters (sky-projected angle, obliquity and size)
  using only a small part of the transit. Substructure in the rings, e.g.  gaps,
  can be recovered if the width of the features ($\delta W$) relative to the
  size of the star is similar to the intrinsic velocity resolution (set by the
  width of the local stellar profile, $\gamma$) relative to the stellar rotation
  velocity ($v$ sin$i$, i.e. $\delta W / R_* \ga v$sin$i$/$\gamma$).

  This opens up a new way to study the ring systems around planets with long
  orbital periods, where observations of the full transit, covering the ingress
  and egress, are not always feasible.

\end{abstract}




\section{Introduction}\label{sect:intro}

Within our solar system, ring systems of varying extent are present around
each of the gas- and ice-giants, the most famous being those of Saturn.
Such rings are not only constrained to giant planets, however.
For example, a thin dense ring has been revealed around the Centaur object
Chariklo \citep[][]{BragaRibasEtAl14}, and further evidence of past ring
structure around Iapetus (a satellite of Saturn) has been
unveiled by the Cassini mission \citep[][]{Ip06}. Beyond the confines
of our solar system, a giant ring
system spanning a diameter of $\sim $0.2 -- 0.8 AU has been discovered
around an object transiting the young Sun-like star J1407 \citep[][]{MamajekEtAl12}.
  $\beta$ Pic b may also represent another planetary system with rings
that transit its host star. Its orbit is aligned closely with the line of
sight \citep[e.g.][]{ChauvinEtAl12, MillarBlanchaerEtAl15, WangEtAl16},
and \cite{LecavelierDesEtangsEtAl95} found a $\sim$5\% fluctuation in the lightcurve
of this system in November 1981. The depth of this event indicates the
presence of a transit of a dust disk or ring structure surrounding the planet
\citep[][]{LecavelierDesEtangsAndVidalMadjar09}.

It therefore seems that planetary ring systems may be relatively common
throughout the Universe. Despite this, many open
questions remain about the physics steering ring formation and evolution.
For example, numerous theories regarding the formation of Saturn's rings
have been put forward.
These include the condensation model \citep[where the rings are from the leftover
remnants of a protosatellite disk --][]{Pollack75}, tidal or
collisional disruption of a small moon \citep[e.g.][]{Roche49, Harris84, CharnozEtAl09a, CharnozEtAl09b} or comet \citep[][]{Dones91}, or that it formed from a
super-massive primordial ring. \citep[For a review, see ][]{CharnozEtAl17}.

Another interesting aspect is the possibility of ring-satellite
interactions.  For instance, Saturn's ring system may have given birth
to several satellites (such as Pandora and Prometheus), and the growth
of these satellites may be quite rapid, over time-scales of a few Myrs
\citep[][]{CharnozEtAl10}. Larger ring systems, such as that present
around J1407b, may spawn more massive moons detectable by transit surveys.
Indeed, a gap within the hypothesised ring system of J1407b is consistent with having been
cleared out by a satellite with a mass of up to 0.8 $M_{\oplus}$
\citep[][]{KenworthyAndMamajek15}.

Identifying and characterising ring systems around exoplanets will
therefore help constrain the physical processes governing the
formation and evolution of rings across a wide range of ages and
environments. Detection of ring gaps may also inform us of the
possible presence of exomoons (some potentially habitable),
providing an additional discovery path to these objects \citep[][]{Kenworthy17}.
Exo-rings may be observed during
transit, where they produce distinct variations in the lightcurve \citep[as
seen in the J1407b system by SuperWASP, ][]{MamajekEtAl12}. In
addition, \cite{OhtaEtAl09} outlined how exo-rings introduce an
additional velocity anomaly while observing the Rossiter McLaughlin
(RM) effect on top of the RM anomaly arising from the planet
itself.

In this paper, we propose a new technique that uses all the information
encoded in the stellar line profile during transit to directly
characterise the ring-system geometry. This has the advantage over other
methods presented in the literature in that it can yield important information
on transiting ring systems using only a small part of the
transit, rather than requiring coverage of the full transit.
This is particularly pertinent given, for example, the expected duration of 
some of these events. In the case of $\beta$ Pic-b, the transit of the
planet/ring system is expected to take $\sim$2 days -- making
full coverage of this long event difficult. Given the predictions
of upcoming transits of $\beta$ Pic-b in 2017/2018
\citep[e.g.][]{LecavelierDesEtangsAndVidalMadjar16, WangEtAl16},
this motivated our development of this new method.

Since $\beta$ Pic is a fast rotator, the velocity gradient
across the star provides an additional handle on the shape of the
rings during transit. Here we present an investigation into the impact
of rings on the stellar line profiles of a $\beta$ Pic-like system.
In Sect.~\ref{sect:model} we  present our model, which we use in
Sect.~\ref{sect:depend} to investigate the dependence of  the line
profile distortions on different parameters of the ring system.  In
Sect.~\ref{sect:recovery} we show (through simulations) how we can
recover the parameters of the ring  system, and we discuss the results
in Sect.~\ref{sect:discuss}. Finally, we present our conclusions in
Sect.~\ref{sect:concl}.

\section{Modelling the line-shape distortions due to transiting ring systems}\label{sect:model}

We model the integrated stellar line profile as a function of velocity,
F$_0$($v$), using a 2d grid.
We assign each pixel [i,j] on the stellar surface an intrinsic line profile,
f$_{ij}$($v$) taking into account the stellar rotation, characterised by
$v$sin$i$, assuming solid body rotation, and an intrinsic line broadening, $\gamma$, which is assumed to be the same for all locations on the star.
In this paper we model the intrinsic line profile as a Gaussian:
\begin{equation}\label{eqn:fij}
f_{ij}(v)=1.-A e^{-\frac{(4 ln(2) )(v-v_{r,ij})^2}{\gamma^2}}
\end{equation}
where A is the maximum strength at the centre of the line, and $v_{r,ij}$ is the velocity of the centre of the line due to stellar rotation at that position.

In addition, we take limb-darkening into account using the quadratic
limb-darkening law \citep[e.g.][]{claret2000}. The total integrated line
profile is calculated as

\begin{equation}\label{eqn:F0}
  F_0(v)= \sum\limits_{i,j}^{star}f_{ij}(v)(1-u_1(1-\mu_{ij})-u_2(1-\mu_{ij})^2  )
\end{equation}
where u$_1$ and u$_2$ are the linear and quadratic limb-darkening coefficients,
respectively, and $\mu_{ij}$ is the cosine of the angle between the line of sight
and the emission at that location. 

The stellar disk was modelled using $\sim$817,000 pixels, giving a radius of
510 pixels. The disk-integrated line profile (determined using
Equation~\ref{eqn:F0}) was calculated at a velocity resolution of
250 m s$^{-1}$ from -250 km s$^{-1}$ to +250 km s$^{-1}$. The planet was
modeled as a fully opaque disk with radius R$_p$, while the rings were
modeled as concentric ellipses, centred on the planet.
The rings were all assumed to be co-planar (i.e. they all have the
same obliquity and position angle (PA) on the sky), and to be circular
when viewed face on. Each ring is described by three parameters, its inner
radius R$_{min,n}$, its outer radius R$_{max,n}$ and its optical depth
$\tau_n$ and are considered to be nested
(i.e. R$_{min,n+1}\gid$R$_{max,n}$). Finally, we assume the rings to
be thin, such that we can specify the optical depth for each ring independent of the inclination.

To model the transit we assume a circular orbit for the planet, with
an orbital period $P$, semi-major axis $a$, and an impact
parameter $b$. Here we define $b$ as the minimum distance between the centre of the stellar disk and the centre of the planet's disk.
For the simulations considered here we also assume spin-orbit alignment, i.e. that the orbital momentum vector of the
planet's orbit is (almost) parallel to the stellar rotation axis. The close alignment of the stellar rotation and orbital axes in the solar system \citep[e.g.][]{Giles2000} is attributed to the formation of the Sun and planets from a single rotating proto-stellar disc that was also initially aligned perpendicular to the solar-rotation axis. 
Studies of short period hot Jupiter systems, however, show a significant spin-orbit misalignment for stars with T$_{eff} \gtrsim $ 6250 K \citep[e.g.][]{FabryckyAndWinn09, MazehEtAl15}, where it is thought that the primordial spin-orbit alignment was disrupted by the process of migration. These results may not
necessarily apply to $\beta$ Pic b-like planets with orbital periods
of several years. Indeed, in the case of $\beta$ Pic b itself, \cite{CurrieEtAl11} report that the planet's orbit is aligned with the flat
outer debris disk. \cite{WatsonEtAl11} and \cite{GreavesEtAl14} showed
that there was no observational evidence for misalignments between
stars and their debris discs, and that the general picture was one of
good star-disk alignment. These studies also included some debris disk host
stars with imaged planetary candidates between 15 and 180 AU that further
suggested planet-disc co-planarity. Thus, the assumption of spin-orbit
alignment in our models appears reasonable when considering long period planets. Furthermore, a small misalignment will have a negligible effect on the shape of the distortions.

Finally, we calculate the position of the planet as a function of time, and determine the line profile observed during transit, F($v$,t) by subtracting the flux under the planet and rings from the full disk-integrated line
profile F$_0$($v$) taken when the planet and rings are completely off
the stellar disk:

\begin{equation}\label{eqn:Fp}
  F(v,t)= F_0(v) - \sum\limits_{n}\sum\limits_{i,j}^{ring_n}f_{ij}(v)(1-e^{-\tau_n})(1-u_1(1-\mu_{ij})-u_2(1-\mu_{ij})^2)
\end{equation}

\section{The dependency of the line profiles on the ring properties}\label{sect:depend}
In the previous section we presented our model for simulating the
expected line-profiles resulting from a transiting planet plus
ring-system.  In this section we investigate the impact that different
parameters have on the observed line profiles. For these simulations
we set the stellar parameters to approximate those of $\beta$
Pic. We assume a stellar rotation of $v$sin$i$=130 km s$^{-1}$, an
intrinsic line width of $\gamma$=20 km s$^{-1}$ FWHM, and V-band limb
darkening coefficients for an effective temperature of 8000 K and
$\log g$=4.0 from~\cite{claret2000}.

We set the planet-to-star radius ratio to R$_p$/R$_*$=0.1 (comparable
to the radius ratio derived from the radius measurement for $\beta$
Pic b from~\citealt{CurrieEtAl13}), and consider a simple ring system with
two rings that have the same opacity, $\tau$. We vary the optical
depth from 0 to 2.5 in steps of 0.125 (note that for Saturn the
optical depth varies from $\sim$0.05 to $\gtrsim$5
\cite[e.g.][]{Colwell09,HedmanAndNicholoson16}).   The inner ring starts at 1.8 R$_p$ and
extends to 2.8 R$_p$ and  the outer ring starts at 3.8 R$_p$ and extends
to 4.8 R$_p$. We note that the rings modelled here are larger than
Saturn's rings, which extend from $\sim$1.2 -
2.3 R$_{Saturn}$,~\citep{allensastrquant}, as we are considering more massive ($\beta$ Pic b like) planets.
To show the effects of
altering the ring parameters more clearly, we also calculated residual
line profiles, F$_{res}$($v$), with respect to the unocculted stellar
profile, F$_0$($v$) \citep[e.g.][]{CeglaEtAl16}.

\begin{equation}
   F_{res}(v,t)=F(v,t)-F_0(v)
\end{equation}

We note that the correct normalisation of F(v,t) is very important, and requires a precise knowledge of the transit lightcurve in order to obtain these residual profiles. For real data, this may not always be possible, however, we show
that the parameters of the ring system can still be recovered by fitting the full profiles (see Sect.~\ref{sect:recovery}).

\begin{figure*}
	\includegraphics[width=\textwidth]{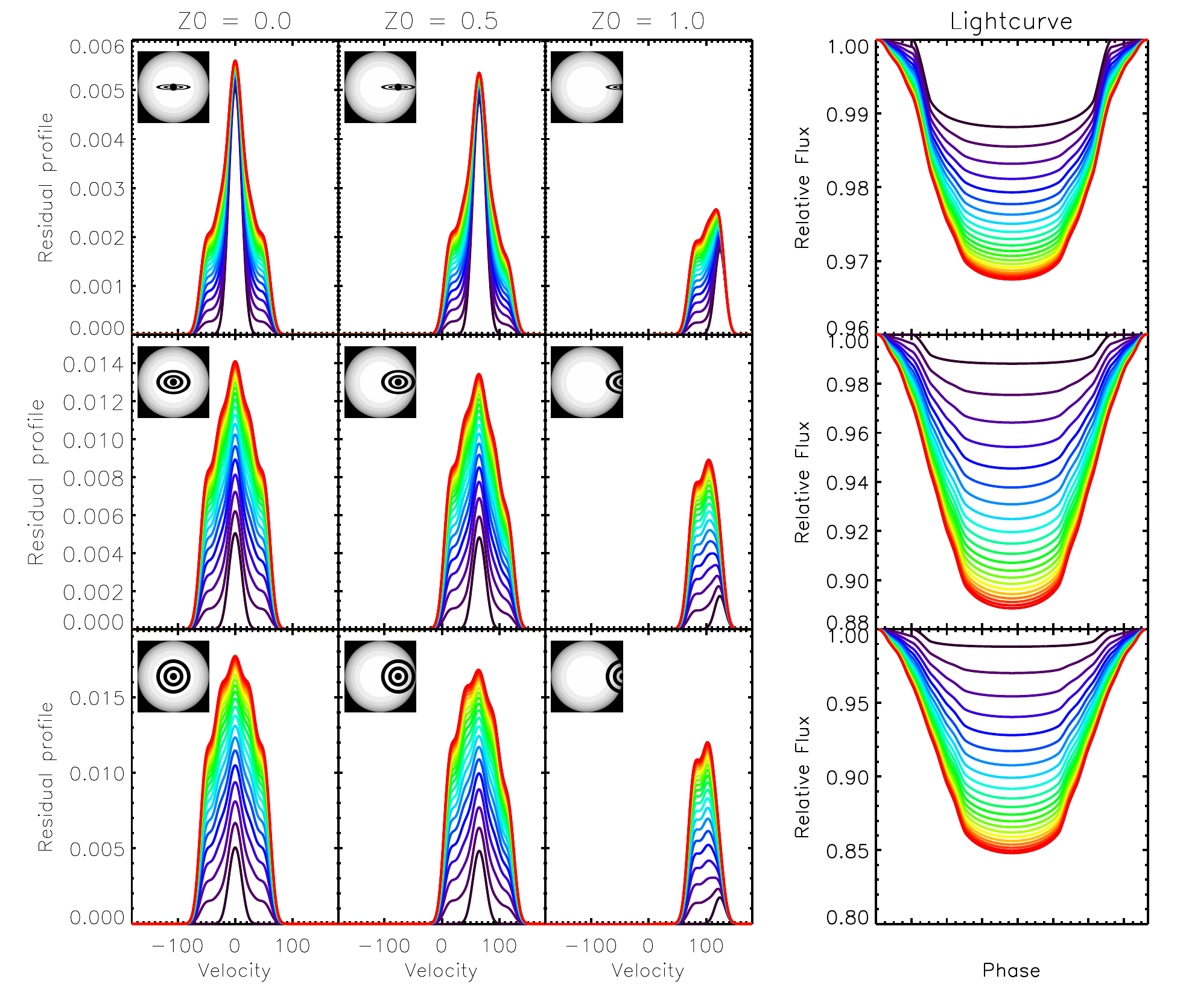}
    \caption{Illustration of the impact of the obliquity of the rings on the observed residual line profiles (left three panels) and lightcurve (right panels). The top row is for an obliquity of 80 degrees, the middle row for an obliquity of 45 degrees and the bottom row for an obliquity of 0 degrees. The residual line profiles are plotted for different distances from mid-transit, $Z_0$, given in units of the stellar radius. Note that, for clarity, we have removed the offset between the individual residual profiles. The colour scale for both the lightcurves and residual line profiles shows the effect of increasing opacity, $\tau$ (Black to red: $\tau$=0 to $\tau$=2.5). For these simulations the position angle on the sky was set to PA=90$\degr$, the impact parameter b=0, and the intrinsic line profile has a FWHM $\gamma$=20 km s$^{-1}$ }
    \label{fig:incl}
\end{figure*}

\begin{figure*}
 	\includegraphics[width=\textwidth]{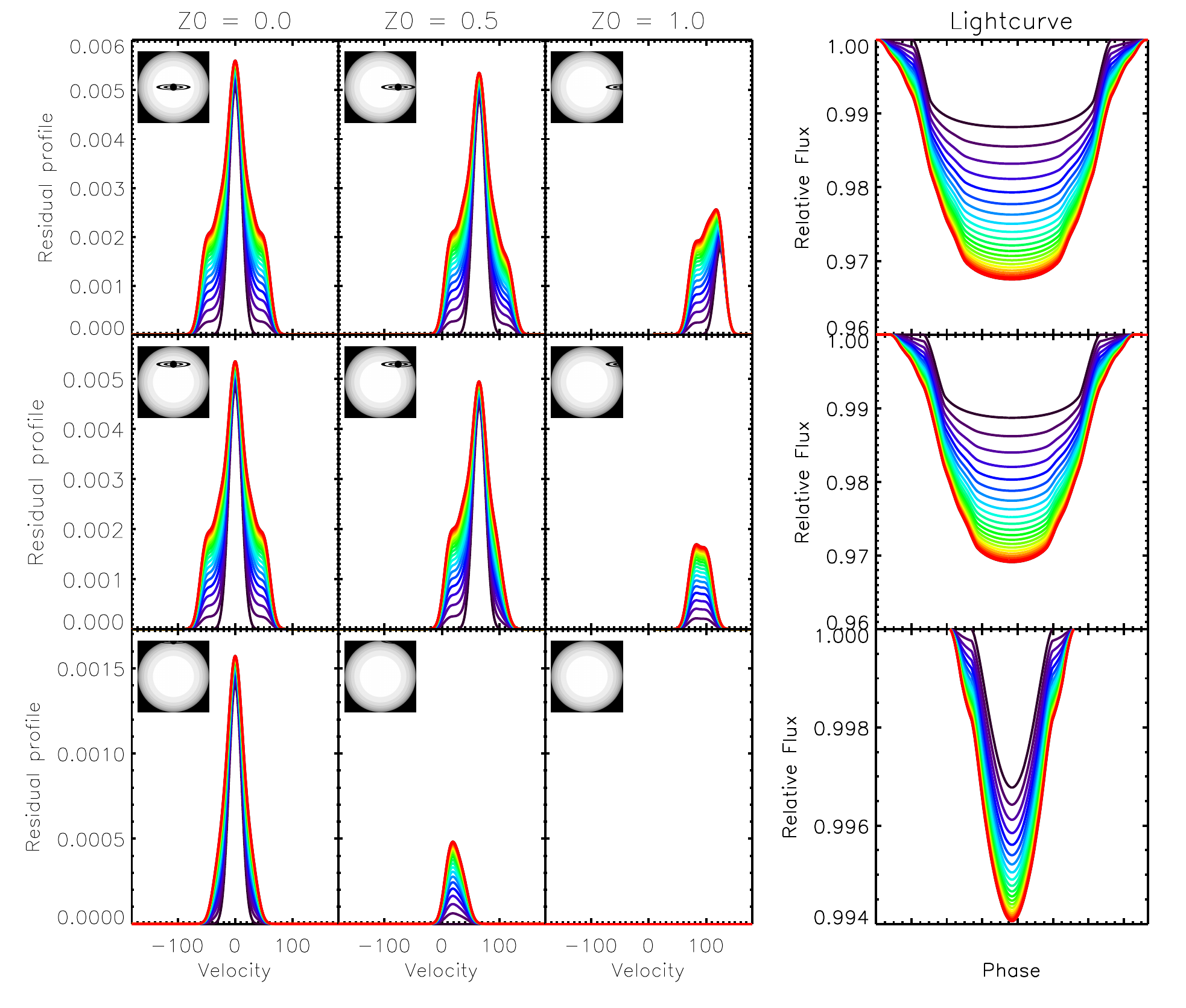}
    \caption{Same as Fig.~\ref{fig:incl}, but now showing the effect of increasing the impact parameter, $b$. From top to bottom we show $b$=0, $b$=0.5 and $b$=1.0. For these simulations the rings were inclined by 80$\degr$ from face-on.}
    \label{fig:impact}
\end{figure*}

\begin{figure*}
	\includegraphics[width=\textwidth]{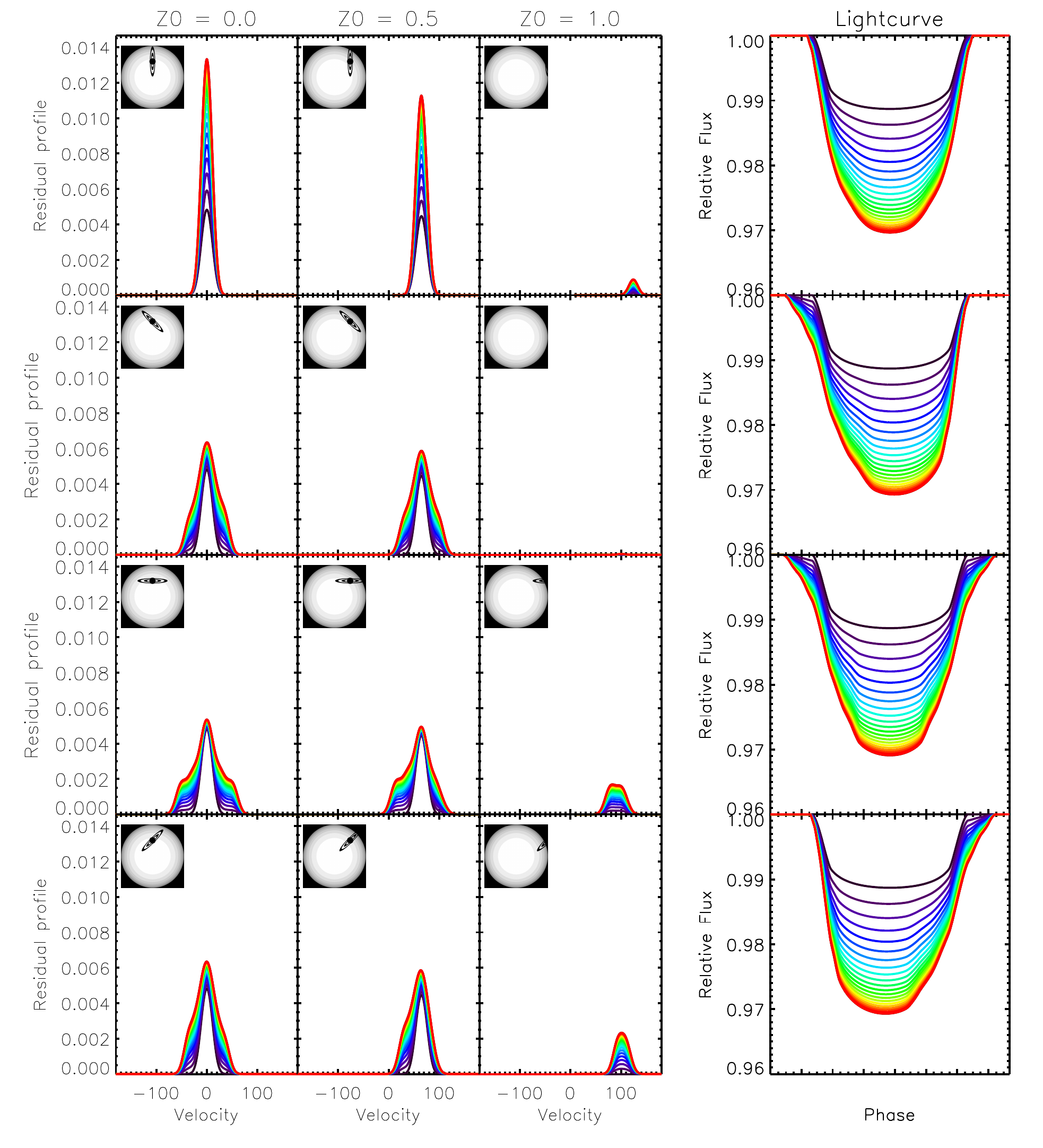}
    \caption{Same as Fig.~\ref{fig:incl}, but now showing the effect of the position angle of the rings. From top to bottom we show PA=0$\degr$, PA=45$\degr$, PA=90$\degr$ and PA=135$\degr$.  For these simulations the rings were inclined by 80$\degr$ from face-on.}
    \label{fig:pa}
\end{figure*}

In Figs.~\ref{fig:incl}~to~\ref{fig:gamma} we show the impact of varying the obliquity of the rings, the impact parameter, $b$, the position angle of the rings on the sky, and the width of the intrinsic line-profile, $\gamma$. We also show the impact of these parameters on the transit lightcurve. A cursory inspection
of the results presented in Figs.~\ref{fig:incl}~to~\ref{fig:gamma}
clearly shows that the presence of a ring system can have a significant
impact on the line profiles -- and their presence can be inferred from a
single `snapshot'. This is in contrast to the transit lightcurves, where the
clearest signal of the presence of a ring system 
occurs during ingress and egress -- requiring good data sampling at those times.

\subsection{Varying the ring parameters}
As can be seen from Fig.~\ref{fig:incl}, for a nearly edge-on ring
system, increasing the opacity does not have a significant impact on
the centre of the line-profile. This is as expected, since the rings
mainly occult areas of the star that are offset in velocity from the
planet's disk. In addition, the gap between the rings is visible as a
flattening of the profile between 40 km s$^{-1}$ and 50 km
s$^{-1}$. This is in line with expectations for a star with
$v$sin$i$=130 km s$^{-1}$ and a ring gap between 0.28 and 0.38 R$_*$,
for which the projection of the ring gap corresponds to an offset from the
centre of the line profile by 36 km s$^{-1}$ (at the inner edge) to
49 km s$^{-1}$ (at the outer edge).

As the planet moves along its orbit, the asymmetry in the profiles
become more apparent. In the first instance this asymmetry is
caused by the gradient of the limb-darkening across the stellar disk,
while closer to egress the asymmetry is caused by the fact that only
part of the planet and rings are occulting the star. When changing the
obliquity towards a face-on ring system, the presence of two rings
separated by a gap becomes more obvious as $\tau$ increases. This can be
identified by the presence of the 2 `shoulders' either side of the centre of
the residual line profiles, and is most clearly seen at mid-transit.

The effect of changing the impact parameter b is shown in
Fig.~\ref{fig:impact}. It is quite evident that, as expected, the
duration of the transit decreases with increasing impact
parameter. The effect on the residual line profiles is minor, except
when the impact parameter is at 1 and only a small portion of the rings
occult the star due to the grazing nature of the transit and nearly edge-on
viewing angle of the rings in this particular simulation.

We show the impact of changing the position angle on the sky in
Fig.~\ref{fig:pa}. When the rings are aligned perpendicular to the
orbit of the planet, the residual line profile remains narrow, but the
amplitude of the residual profile increases significantly with
increasing $\tau$. However, since there is no velocity gradient along
the major axis of the rings, the ring gap is not clearly
detectable. When the rings are misaligned with the planet's orbit
(e.g. for PA=$\pm$45$\degr$), the transit lightcurve becomes highly
asymmetric. The line profiles are also asymmetric, especially for
z$_0\la$0 for PA$>$0 and z$_0\ga$0 for PA$<$0. For these position
angles the rings occult parts of the star with very different surface
brightnesses, while on the other side, the gradient in the limb-darkening
from one end of the rings to the other is reduced.

\subsection{The effect of the intrinsic line-profile width}
\begin{figure*}
	\includegraphics[width=\textwidth]{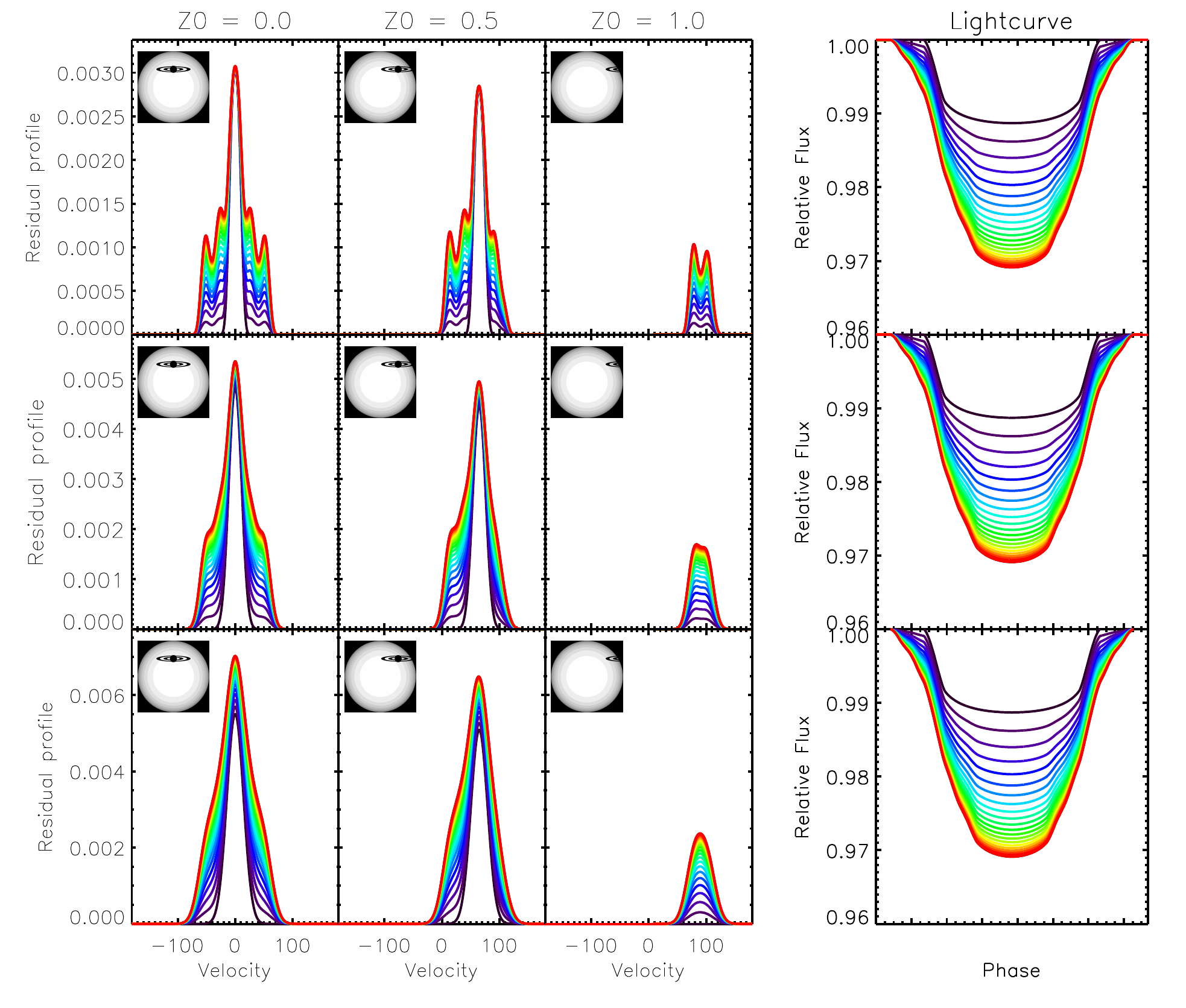}
    \caption{Same as Fig.~\ref{fig:incl}, but now showing the effect of increasing the impact of the width of the intrinsic stellar line profile. From top to bottom we show the residual profiles for a FWHM of $\gamma$=10 km s$^{-1}$, $\gamma$=20 km s$^{-1}$, and $\gamma$=30 km s$^{-1}$. For these simulations the rings were inclined by 80$\degr$ from face-on. }
    \label{fig:gamma}
\end{figure*}

The intrinsic line profile width, $\gamma$, sets the fundamental limit
that defines how well features can be resolved. In
Fig.~\ref{fig:gamma} we show the impact of changing the FWHM of the
intrinsic line profile from 10 km s$^{-1}$ to 30 km s$^{-1}$. It is
clear that for a lower $\gamma$ it is easier to resolve features in
the rings, and the two gaps can be clearly seen. However, as $\gamma$
is increased to 30 km s$^{-1}$ it becomes harder to discern the
presence of a clear ring gap. This implies that fast rotating stars
with a narrow intrinsic line-profile are preferable for this type of
observation, and that care should be taken with the selection of the
wavelength region to be observed such that lines with a lower intrinsic broadening are targeted. 

\section{Recovering the parameters of the rings from simulated observations}\label{sect:recovery}
In the previous section we showed the impact of different parameters
on the observed line-profile changes. However, since in reality it
will be very difficult to obtain absolute spectra, getting direct
measurements of line-profile variations will be difficult to
achieve. It is therefore more useful to test how well the parameters
can be recovered when attempting to fit the continuum normalised
line-profiles that are typical of high-resolution spectroscopic
observations. To do this we generated a set
of transit observations of different duration and with different
properties of the rings to which we add noise, and that we
subsequently fit using a Markov Chain Monte Carlo (MCMC) method.

To simulate the observations, we assume that the planet moves on a
circular orbit, with a semi-major axis of 8.2 AU and an orbital period
of 18 years, similar to the short period case for $\beta$ Pic from
\cite{LecavelierDesEtangsAndVidalMadjar16}. For these simulations we adopted
an impact parameter of $b$=0.5. We have also taken into account
the resolution of the instrument, assuming a value of R$\sim$100,000.
Since our simulated model profiles are oversampled at a resolution of
250 m s$^{-1}$, we first convolve the simulated line profiles with a
Gaussian at the instrumental
resolution, and then rebin the convolved profiles to a grid of 1 km
s$^{-1}$, which is approximately the sampling used by most
R$\sim$100,000 spectrographs.  We also assume that the noise in each pixel
is Gaussian with $\sigma$=0.0014 (SNR$\sim$700), which, for a star of
the brightness of $\beta$ Pic, should be achievable with a 1.5 minute
cadence when using a high-resolution echelle spectrograph on an
8m-class telescope.

For each set of planet and ring parameters, we run two sets of
simulations to test the impact of the timing of the observations on
the ability to recover the properties of the ring. The first set is
taken just after mid-transit, while the second set is taken midway
through egress. For each of the two sets we simulate a single
observation, a single block of 10 sequential observations (lasting 15
minutes in total), and finally a single block of 100 observations
(lasting 2.5 hours in total), again all exposures are assumed to be taken sequentially.

For the simulations we vary the position angle between 0$\degr$ and
135$\degr$ in steps of 45$\degr$ for two different obliquities,
10$\degr$ from edge on and at 45$\degr$ from edge on. As before, we
set R$_p$/R$_*$=0.1, but now simulate a slightly smaller system with 3
rings without gaps extending from 1.3-4.3 R$p$. The first ring starts at
1.3 R$_p$ and ends at 2.3 R$_p$ and has an optical depth of
$\tau$=1. The middle ring has $\tau$=0.5 and ends at 3.3R$_p$. The
final ring ends at 4.3R$_p$ and has an optical depth of $\tau$=1.

After generating the individual simulations we fit each of them
using a model with a single ring. To facilitate the fitting over a
large parameter range we used a simple MCMC fit consisting of chains of
50,000 steps.

\section{Discussion}\label{sect:discuss}

\begin{figure*}
	\includegraphics[width=\textwidth]{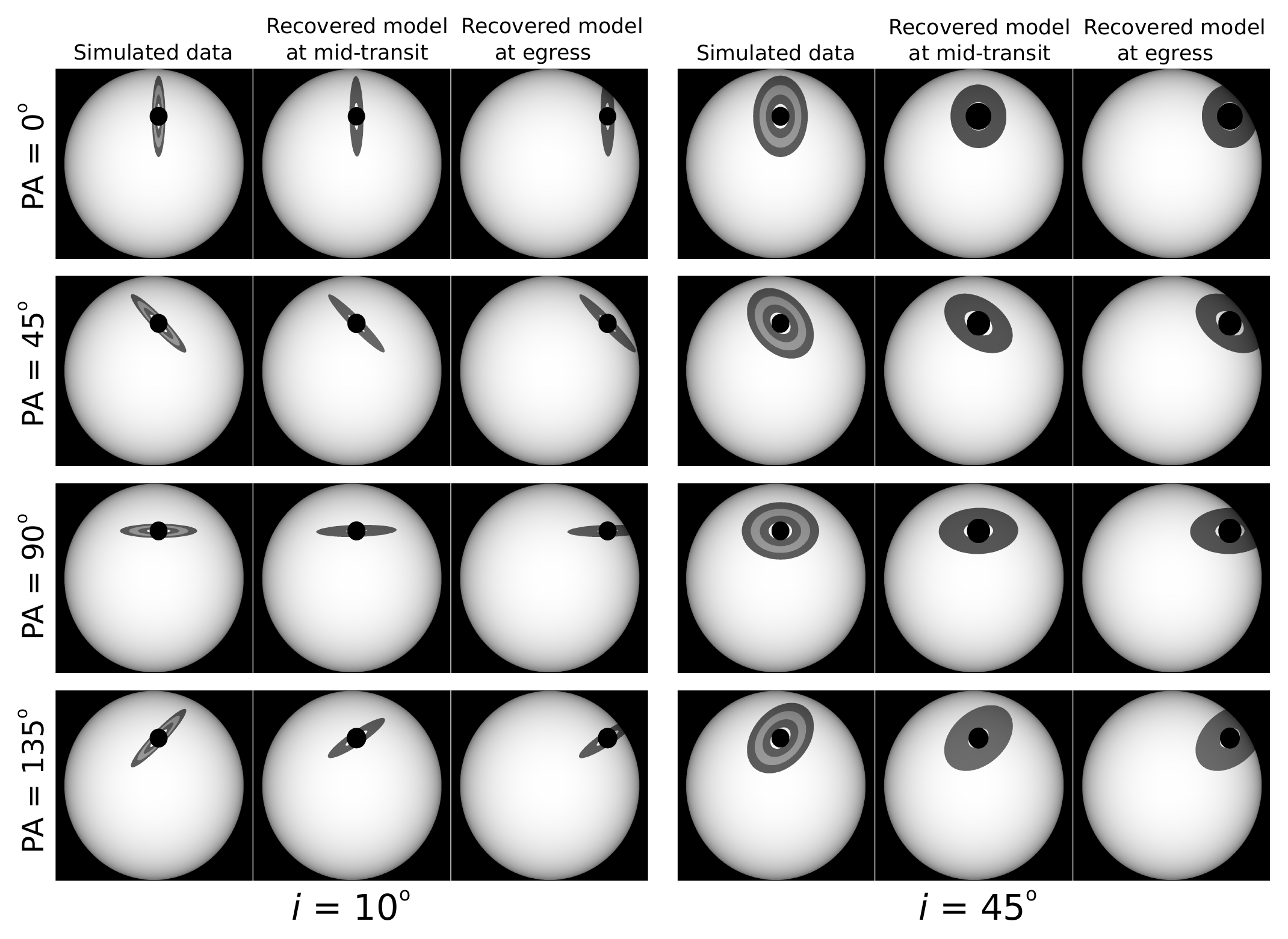}
    \caption{Examples of the simulated and recovered ring systems. For each set of three images, the left image shows the simulations just after mid-transit, the middle shows the best-fit model at mid-transit and the right panel shows the best fit model close to egress. All the fits were done to the simulations covering 2.5 hrs of observing time (100 frames). The left columns shows simulations for a rins with an obliquity of 10$\degr$ from edge on, while the right column is for an obliquity of 45$\degr$. From top to bottom the simulations are for a PA of 0$\degr$, 45$\degr$, 90$\degr$ \& 135$\degr$, respectively. }
    \label{fig:recov_example}
\end{figure*}

\begin{figure*}
	\includegraphics[width=\textwidth]{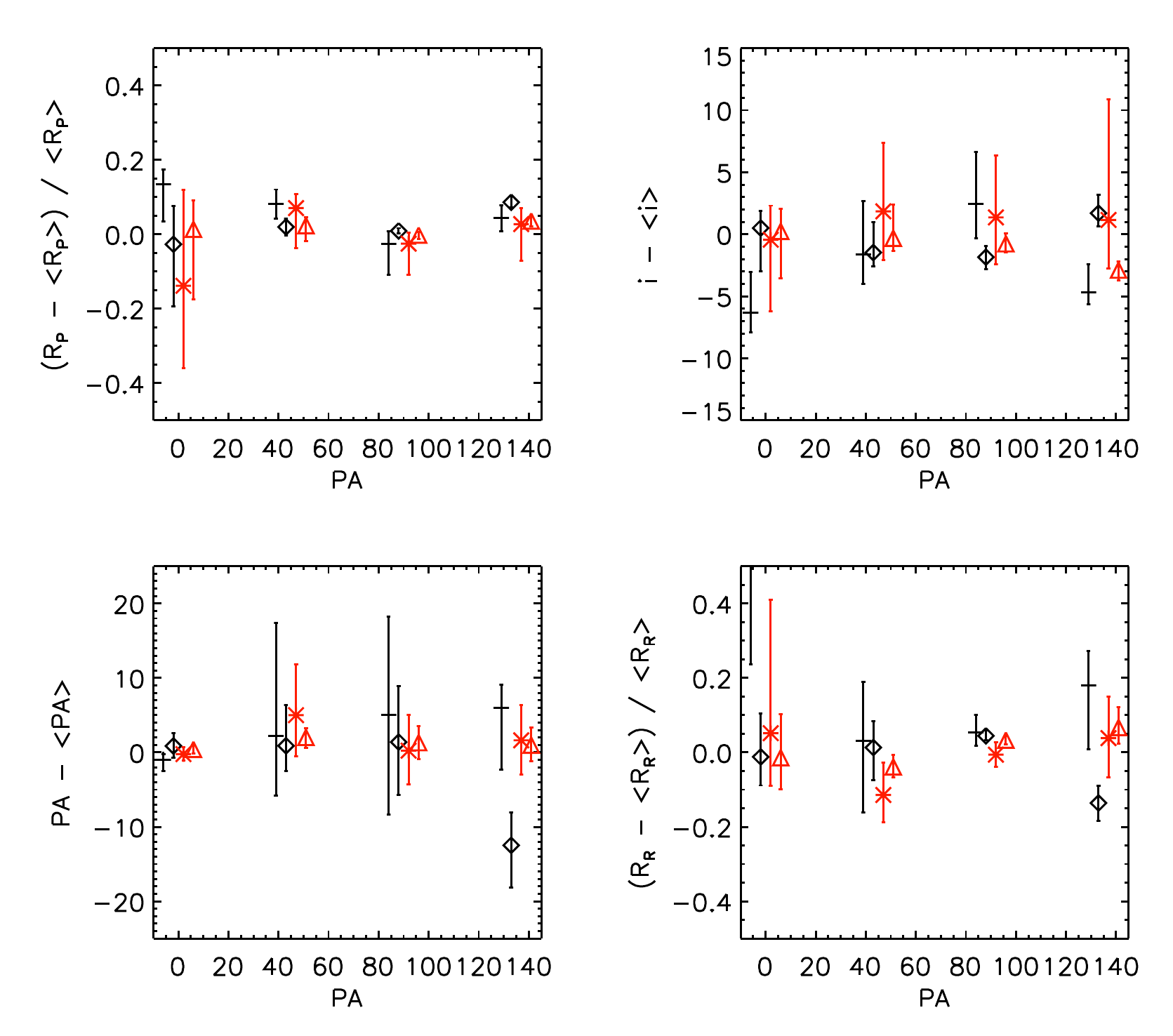}
    \caption{Differences between the recovered and original parameters for our simulations for an obliquity of 10$\degr$. Top left: Relative difference in the radius of the planet (R$_p$). Top right: Difference in the recovered obliquity of the rings ($i$). Bottom left: Difference in the position angle of the rings (PA). Bottom right: Relative difference in the radius of the rings (R$_R$). The black points show the results for simulated observations near mid-transit, while the red points show the results for the simulated observations near egress. The plus and star symbols are for 15 minutes of observations (10 frames), while open diamonds and triangles are for observations lasting 150 minutes (100 frames). Note that the points have been offset in PA for clarity. }
    \label{fig:recov10}
\end{figure*}

\begin{figure*}
	\includegraphics[width=\textwidth]{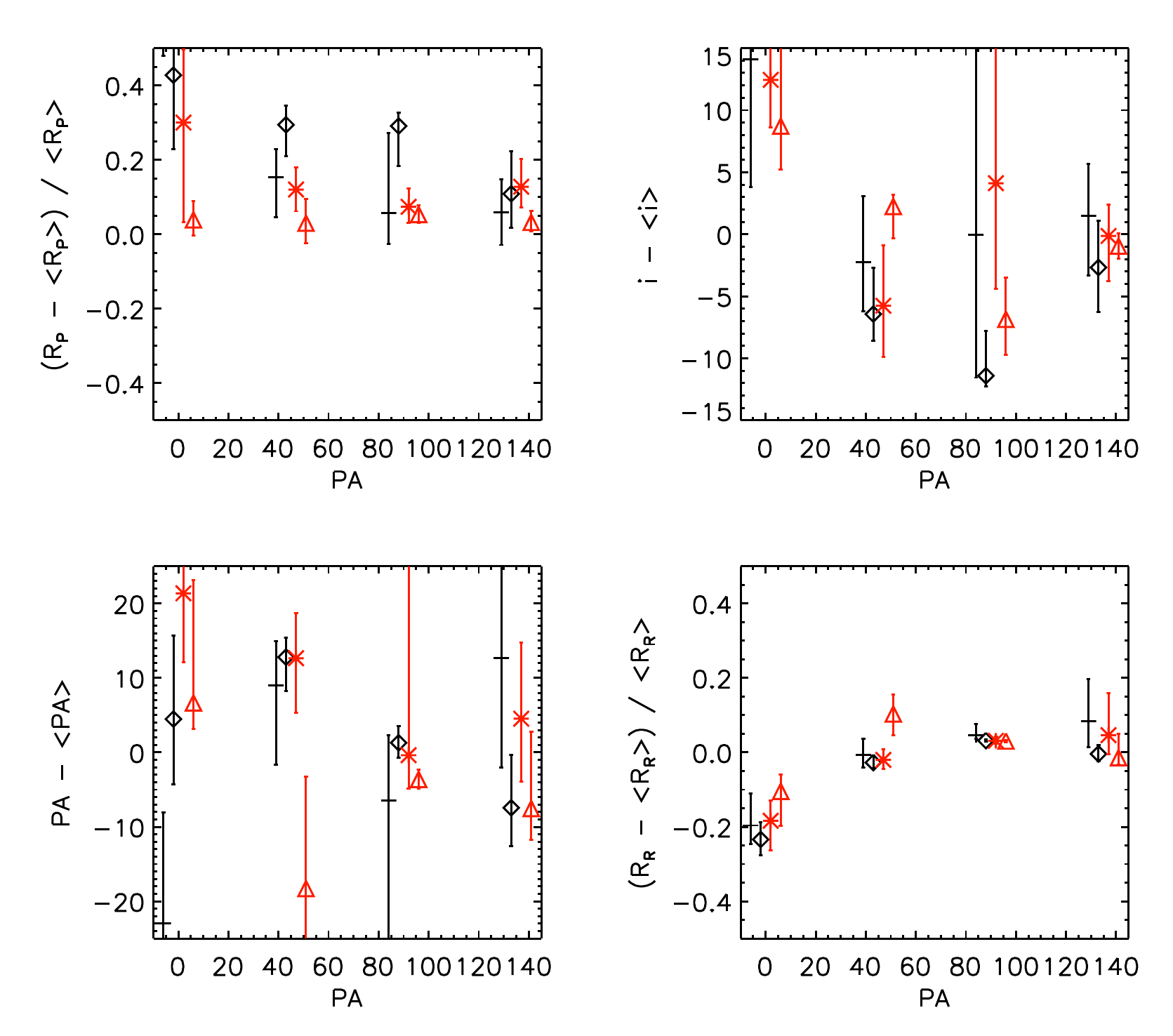}
    \caption{ Same as for Fig.~\ref{fig:recov10}, but now for an obliquity of 45$\degr$. }
    \label{fig:recov45}
\end{figure*}

The snap-shot simulations consisting of a single frame did not provide reasonable constraints on the recovered parameters, and we therefore focus on the simulations consisting of blocks of 10 and 100 frames (15 minutes and 150 minutes).

From our simulations and subsequent fits we find that, in general, we
can obtain reasonable constraints on the main system parameters
(obliquity, position angle, and outer ring radius).
Fig.~\ref{fig:recov_example} shows a visual comparison between the
simulations and the recovered models for the simulated data sets
lasting 150 minutes (100 frames). By eye, it is already quite clear
that in most cases we can obtain useful constraints on the size,
position angle and obliquity of the ring system. The exception is for
an obliquity of 45$\degr$ and a position angle of 0$\degr$ (top right
images in Fig.~\ref{fig:recov_example}), where the recovered
parameters result in a ring that is clearly more compact and
circular. We attribute this to the fact that at an obliquity of
45$\degr$ the rings are already more circular and, to complicate
matters further, at a PA of 0$\degr$ the projected major axis of the
rings is aligned with the stellar rotation axis, leading to a
degeneracy between the opacity and extent of the rings parallel to the
projected stellar rotation axis.

From the images in Fig.~\ref{fig:recov_example} it is also clear that
we are not able to fit the interior gap properly. This is not
surprising, as in our input models the gap is only 0.03R$_*$
wide. This corresponds to an extent in velocity of $\Delta v\sim$4 km
s$^{-1}$), approximately four times lower than the intrinsic
resolution of the line profiles. When fitting, it is therefore
possible to make a trade-off between the size of the gap (location of
the inner edge of the ring) and the radius of the planet.

A more qualitative analysis of our ability to recover the ring
parameters is presented in Figs.~\ref{fig:recov10} and \ref{fig:recov45} , where we show the fractional
differences between the input parameters and the best fit parameters
from our MCMC analysis. Again, it is clear that we can recover most
parameters quite well. The exception is at PA=0$\degr$, when the
major-axis of the rings is parallel to the projected rotation axis of
the star, and the rings occult a very limited range in projected
rotational velocities across the stellar disk.  For an obliquity of 45$\degr$ (Fig.\ref{fig:recov45}),the parameters are less well constrained than for the case of an obliquity of 10$\degr$. This can be understood by the fact that for a an increased obliquity the rings appear more symmetrical, and changes in the orientation of the rings have a relatively smaller impact  on the distortions in the stellar line-profile.

Our simulations indicate that we can recover several fundamental
parameters of the ring system, including the outer radius of the
rings, the position angle of the disk and the obliquity of the rings,
without the need to observe the entire transit. This opens up the
possibility to study ring systems for long period planets, where the
transit duration exceeds the length of a single night. As expected, the
simulations for 100 frames (2.5 hours of observation) provide a more robust
constraint on the parameters than the simulations for 15 minutes of observations.

\section{Conclusions}\label{sect:concl}

Using a simple model, we have shown that the added dimension of the
stellar rotation allows us to directly determine the properties of
rings around exoplanets that transit fast-rotating stars. We have also
shown that this type of observation does not require us to observe
the entire transit, making them particularly useful for planets at
large orbital separations. However, we note that for higher obliquities, the properties of the ring system are less well constrained. This is due to the fact that for these systems the rings appear more symmetrical. Furthermore, for rings that have their projected semi-major axis parallel to the stellar axis of rotation, the reduced amount of velocities covered by the planet results reduces the constraints on the properties of the rings.
We have shown the impact of different ring
parameters on the distortion of the line profile, and demonstrated that
the effects of individual rings become more obvious for narrower
intrinsic line-profiles.

\section*{Acknowledgements}

EdM was in part funded by the Michael West Fellowship.
C.A.W. acknowledges support by STFC grant ST/P000312/1.
%




\bibliographystyle{mnras}
\bibliography{rm_rings} 

\begin{thebibliography}{}
\makeatletter
\relax
\def\mn@urlcharsother{\let\do\@makeother \do\$\do\&\do\#\do\^\do\_\do\%\do\~}
\def\mn@doi{\begingroup\mn@urlcharsother \@ifnextchar [ {\mn@doi@}
  {\mn@doi@[]}}
\def\mn@doi@[#1]#2{\def\@tempa{#1}\ifx\@tempa\@empty \href
  {http://dx.doi.org/#2} {doi:#2}\else \href {http://dx.doi.org/#2} {#1}\fi
  \endgroup}
\def\mn@eprint#1#2{\mn@eprint@#1:#2::\@nil}
\def\mn@eprint@arXiv#1{\href {http://arxiv.org/abs/#1} {{\tt arXiv:#1}}}
\def\mn@eprint@dblp#1{\href {http://dblp.uni-trier.de/rec/bibtex/#1.xml}
  {dblp:#1}}
\def\mn@eprint@#1:#2:#3:#4\@nil{\def\@tempa {#1}\def\@tempb {#2}\def\@tempc
  {#3}\ifx \@tempc \@empty \let \@tempc \@tempb \let \@tempb \@tempa \fi \ifx
  \@tempb \@empty \def\@tempb {arXiv}\fi \@ifundefined
  {mn@eprint@\@tempb}{\@tempb:\@tempc}{\expandafter \expandafter \csname
  mn@eprint@\@tempb\endcsname \expandafter{\@tempc}}}

\bibitem[\protect\citeauthoryear{{Braga-Ribas} et~al.,}{{Braga-Ribas}
  et~al.}{2014}]{BragaRibasEtAl14}
{Braga-Ribas} F.,  et~al., 2014, \mn@doi [\nat] {10.1038/nature13155}, \href
  {http://adsabs.harvard.edu/abs/2014Natur.508...72B} {508, 72}

\bibitem[\protect\citeauthoryear{{Cegla}, {Lovis}, {Bourrier}, {Beeck},
  {Watson}  \& {Pepe}}{{Cegla} et~al.}{2016}]{CeglaEtAl16}
{Cegla} H.~M.,  {Lovis} C.,  {Bourrier} V.,  {Beeck} B.,  {Watson} C.~A.,
  {Pepe} F.,  2016, \mn@doi [\aap] {10.1051/0004-6361/201527794}, \href
  {http://adsabs.harvard.edu/abs/2016A%26A...588A.127C} {588, A127}

\bibitem[\protect\citeauthoryear{{Charnoz}}{{Charnoz}}{2009}]{CharnozEtAl09b}
{Charnoz} S.,  2009, \mn@doi [\icarus] {10.1016/j.icarus.2008.12.036}, \href
  {http://adsabs.harvard.edu/abs/2009Icar..201..191C} {201, 191}

\bibitem[\protect\citeauthoryear{{Charnoz}, {Morbidelli}, {Dones}  \&
  {Salmon}}{{Charnoz} et~al.}{2009}]{CharnozEtAl09a}
{Charnoz} S.,  {Morbidelli} A.,  {Dones} L.,   {Salmon} J.,  2009, \mn@doi
  [\icarus] {10.1016/j.icarus.2008.10.019}, \href
  {http://adsabs.harvard.edu/abs/2009Icar..199..413C} {199, 413}

\bibitem[\protect\citeauthoryear{{Charnoz}, {Salmon}  \& {Crida}}{{Charnoz}
  et~al.}{2010}]{CharnozEtAl10}
{Charnoz} S.,  {Salmon} J.,   {Crida} A.,  2010, \mn@doi [\nat]
  {10.1038/nature09096}, \href
  {http://adsabs.harvard.edu/abs/2010Natur.465..752C} {465, 752}

\bibitem[\protect\citeauthoryear{{Charnoz}, {Canup}, {Crida}  \&
  {Dones}}{{Charnoz} et~al.}{2017}]{CharnozEtAl17}
{Charnoz} S.,  {Canup} R.~M.,  {Crida} A.,   {Dones} L.,  2017, preprint, \href
  {http://adsabs.harvard.edu/abs/2017arXiv170309741C} {} (\mn@eprint {arXiv}
  {1703.09741})

\bibitem[\protect\citeauthoryear{{Chauvin} et~al.,}{{Chauvin}
  et~al.}{2012}]{ChauvinEtAl12}
{Chauvin} G.,  et~al., 2012, \mn@doi [\aap] {10.1051/0004-6361/201118346},
  \href {http://adsabs.harvard.edu/abs/2012A%26A...542A..41C} {542, A41}

\bibitem[\protect\citeauthoryear{{Claret}}{{Claret}}{2000}]{claret2000}
{Claret} A.,  2000, \aap, \href
  {http://adsabs.harvard.edu/abs/2000A%26A...363.1081C} {363, 1081}

\bibitem[\protect\citeauthoryear{{Colwell}, {Nicholson}, {Tiscareno}, {Murray},
  {French}  \& {Marouf}}{{Colwell} et~al.}{2009}]{Colwell09}
{Colwell} J.~E.,  {Nicholson} P.~D.,  {Tiscareno} M.~S.,  {Murray} C.~D.,
  {French} R.~G.,   {Marouf} E.~A.,  2009, {The Structure of Saturn's Rings}.
p.~375, \mn@doi{10.1007/978-1-4020-9217-6_13}

\bibitem[\protect\citeauthoryear{{Cox}}{{Cox}}{2000}]{allensastrquant}
{Cox} A.~N.,  2000, {Allen's astrophysical quantities}

\bibitem[\protect\citeauthoryear{{Currie}, {Thalmann}, {Matsumura},
  {Madhusudhan}, {Burrows}  \& {Kuchner}}{{Currie} et~al.}{2011}]{CurrieEtAl11}
{Currie} T.,  {Thalmann} C.,  {Matsumura} S.,  {Madhusudhan} N.,  {Burrows} A.,
    {Kuchner} M.,  2011, \mn@doi [\apjl] {10.1088/2041-8205/736/2/L33}, \href
  {http://adsabs.harvard.edu/abs/2011ApJ...736L..33C} {736, L33}

\bibitem[\protect\citeauthoryear{{Currie} et~al.,}{{Currie}
  et~al.}{2013}]{CurrieEtAl13}
{Currie} T.,  et~al., 2013, \mn@doi [\apj] {10.1088/0004-637X/776/1/15}, \href
  {http://adsabs.harvard.edu/abs/2013ApJ...776...15C} {776, 15}

\bibitem[\protect\citeauthoryear{{Dones}}{{Dones}}{1991}]{Dones91}
{Dones} L.,  1991, \mn@doi [\icarus] {10.1016/0019-1035(91)90045-U}, \href
  {http://adsabs.harvard.edu/abs/1991Icar...92..194D} {92, 194}

\bibitem[\protect\citeauthoryear{{Fabrycky} \& {Winn}}{{Fabrycky} \&
  {Winn}}{2009}]{FabryckyAndWinn09}
{Fabrycky} D.~C.,  {Winn} J.~N.,  2009, \mn@doi [\apj]
  {10.1088/0004-637X/696/2/1230}, \href
  {http://adsabs.harvard.edu/abs/2009ApJ...696.1230F} {696, 1230}

\bibitem[\protect\citeauthoryear{{Giles}}{{Giles}}{2000}]{Giles2000}
{Giles} P.~M.,  2000, PhD thesis, STANFORD UNIVERSITY

\bibitem[\protect\citeauthoryear{{Greaves} et~al.,}{{Greaves}
  et~al.}{2014}]{GreavesEtAl14}
{Greaves} J.~S.,  et~al., 2014, \mn@doi [\mnras] {10.1093/mnrasl/slt153}, \href
  {http://adsabs.harvard.edu/abs/2014MNRAS.438L..31G} {438, L31}

\bibitem[\protect\citeauthoryear{{Harris}}{{Harris}}{1984}]{Harris84}
{Harris} A.~W.,  1984, in {Greenberg} R.,  {Brahic} A.,  eds, IAU Colloq. 75:
  Planetary Rings. pp 641--659

\bibitem[\protect\citeauthoryear{{Hedman} \& {Nicholson}}{{Hedman} \&
  {Nicholson}}{2016}]{HedmanAndNicholoson16}
{Hedman} M.~M.,  {Nicholson} P.~D.,  2016, \mn@doi [\icarus]
  {10.1016/j.icarus.2016.01.007}, \href
  {http://adsabs.harvard.edu/abs/2016Icar..279..109H} {279, 109}

\bibitem[\protect\citeauthoryear{{Ip}}{{Ip}}{2006}]{Ip06}
{Ip} W.-H.,  2006, \mn@doi [\grl] {10.1029/2005GL025386}, \href
  {http://adsabs.harvard.edu/abs/2006GeoRL..3316203I} {33, L16203}

\bibitem[\protect\citeauthoryear{{Kenworthy}}{{Kenworthy}}{2017}]{Kenworthy17}
{Kenworthy} M.,  2017, \mn@doi [Nature Astronomy] {10.1038/s41550-017-0099},
  \href {http://adsabs.harvard.edu/abs/2017NatAs...1E..99K} {1, 0099}

\bibitem[\protect\citeauthoryear{{Kenworthy} \& {Mamajek}}{{Kenworthy} \&
  {Mamajek}}{2015}]{KenworthyAndMamajek15}
{Kenworthy} M.~A.,  {Mamajek} E.~E.,  2015, \mn@doi [\apj]
  {10.1088/0004-637X/800/2/126}, \href
  {http://adsabs.harvard.edu/abs/2015ApJ...800..126K} {800, 126}

\bibitem[\protect\citeauthoryear{{Lecavelier Des Etangs} \&
  {Vidal-Madjar}}{{Lecavelier Des Etangs} \&
  {Vidal-Madjar}}{2009}]{LecavelierDesEtangsAndVidalMadjar09}
{Lecavelier Des Etangs} A.,  {Vidal-Madjar} A.,  2009, \mn@doi [\aap]
  {10.1051/0004-6361/200811528}, \href
  {http://adsabs.harvard.edu/abs/2009A%26A...497..557L} {497, 557}

\bibitem[\protect\citeauthoryear{{Lecavelier Des Etangs}, {Deleuil},
  {Vidal-Madjar}, {Ferlet}, {Nitschelm}, {Nicolet}  \&
  {Lagrange-Henri}}{{Lecavelier Des Etangs}
  et~al.}{1995}]{LecavelierDesEtangsEtAl95}
{Lecavelier Des Etangs} A.,  {Deleuil} M.,  {Vidal-Madjar} A.,  {Ferlet} R.,
  {Nitschelm} C.,  {Nicolet} B.,   {Lagrange-Henri} A.~M.,  1995, \aap, \href
  {http://adsabs.harvard.edu/abs/1995A%26A...299..557L} {299, 557}

\bibitem[\protect\citeauthoryear{{Lecavelier des Etangs} \&
  {Vidal-Madjar}}{{Lecavelier des Etangs} \&
  {Vidal-Madjar}}{2016}]{LecavelierDesEtangsAndVidalMadjar16}
{Lecavelier des Etangs} A.,  {Vidal-Madjar} A.,  2016, \mn@doi [\aap]
  {10.1051/0004-6361/201527631}, \href
  {http://adsabs.harvard.edu/abs/2016A%26A...588A..60L} {588, A60}

\bibitem[\protect\citeauthoryear{{Mamajek}, {Quillen}, {Pecaut}, {Moolekamp},
  {Scott}, {Kenworthy}, {Collier Cameron}  \& {Parley}}{{Mamajek}
  et~al.}{2012}]{MamajekEtAl12}
{Mamajek} E.~E.,  {Quillen} A.~C.,  {Pecaut} M.~J.,  {Moolekamp} F.,  {Scott}
  E.~L.,  {Kenworthy} M.~A.,  {Collier Cameron} A.,   {Parley} N.~R.,  2012,
  \mn@doi [\aj] {10.1088/0004-6256/143/3/72}, \href
  {http://adsabs.harvard.edu/abs/2012AJ....143...72M} {143, 72}

\bibitem[\protect\citeauthoryear{{Mazeh}, {Perets}, {McQuillan}  \&
  {Goldstein}}{{Mazeh} et~al.}{2015}]{MazehEtAl15}
{Mazeh} T.,  {Perets} H.~B.,  {McQuillan} A.,   {Goldstein} E.~S.,  2015,
  \mn@doi [\apj] {10.1088/0004-637X/801/1/3}, \href
  {http://adsabs.harvard.edu/abs/2015ApJ...801....3M} {801, 3}

\bibitem[\protect\citeauthoryear{{Millar-Blanchaer} et~al.,}{{Millar-Blanchaer}
  et~al.}{2015}]{MillarBlanchaerEtAl15}
{Millar-Blanchaer} M.~A.,  et~al., 2015, \mn@doi [\apj]
  {10.1088/0004-637X/811/1/18}, \href
  {http://adsabs.harvard.edu/abs/2015ApJ...811...18M} {811, 18}

\bibitem[\protect\citeauthoryear{{Ohta}, {Taruya}  \& {Suto}}{{Ohta}
  et~al.}{2009}]{OhtaEtAl09}
{Ohta} Y.,  {Taruya} A.,   {Suto} Y.,  2009, \mn@doi [\apj]
  {10.1088/0004-637X/690/1/1}, \href
  {http://adsabs.harvard.edu/abs/2009ApJ...690....1O} {690, 1}

\bibitem[\protect\citeauthoryear{{Pollack}}{{Pollack}}{1975}]{Pollack75}
{Pollack} J.~B.,  1975, \mn@doi [\ssr] {10.1007/BF00350197}, \href
  {http://adsabs.harvard.edu/abs/1975SSRv...18....3P} {18, 3}

\bibitem[\protect\citeauthoryear{{Roche}}{{Roche}}{1849}]{Roche49}
{Roche} E.,  1849, M\'emoire de la section des sciences, Acad\'emie des
  sciences et des lettres de Montpellier, 1, 243

\bibitem[\protect\citeauthoryear{{Wang} et~al.,}{{Wang}
  et~al.}{2016}]{WangEtAl16}
{Wang} J.~J.,  et~al., 2016, \mn@doi [\aj] {10.3847/0004-6256/152/4/97}, \href
  {http://adsabs.harvard.edu/abs/2016AJ....152...97W} {152, 97}

\bibitem[\protect\citeauthoryear{{Watson}, {Littlefair}, {Diamond}, {Collier
  Cameron}, {Fitzsimmons}, {Simpson}, {Moulds}  \& {Pollacco}}{{Watson}
  et~al.}{2011}]{WatsonEtAl11}
{Watson} C.~A.,  {Littlefair} S.~P.,  {Diamond} C.,  {Collier Cameron} A.,
  {Fitzsimmons} A.,  {Simpson} E.,  {Moulds} V.,   {Pollacco} D.,  2011,
  \mn@doi [\mnras] {10.1111/j.1745-3933.2011.01036.x}, \href
  {http://adsabs.harvard.edu/abs/2011MNRAS.413L..71W} {413, L71}

\makeatother
\end{thebibliography}

\bsp	
\label{lastpage}
\end{document}